\documentclass[12pt]{article}
\usepackage{geometry} % see geometry.pdf on how to lay out the page. There's lots.
\usepackage{graphicx}
\title{The Metaphysics of D-CTCs: On the Underlying Assumptions of Deutsch's Quantum Solution to the Paradoxes of Time Travel}
\author{Lucas Dunlap}
\date{} % delete this line to display the current date

\begin{document}

\maketitle

%%%%INTRODUCTION

\begin{abstract} I argue that Deutsch's model for the behavior of systems traveling around closed timelike curves (CTCs) relies implicitly on a substantive metaphysical assumption. Deutsch is employing a version of quantum theory with a significantly supplemented ontology of parallel existent worlds, which differ in kind from the many worlds of the Everett interpretation. Standard Everett does not support the existence of multiple identical copies of the world, which the D-CTC model requires. This has been obscured because he often refers to the branching structure of Everett as a ÒmultiverseÓ, and describes quantum interference by reference to parallel interacting definite worlds. But he admits that this is only an approximation to Everett. The D-CTC model, however, relies crucially on the existence of a multiverse of parallel interacting worlds. Since his model is supplemented by structures that go significantly beyond quantum theory, and play an ineliminable role in its predictions and explanations, it does not represent a quantum solution to the paradoxes of time travel.

\end{abstract}

\section{Introduction}

The possible existence of a closed timelike curve---a path in spacetime that takes a traveler to his own past---gives rise to the possibility of serious paradoxes. The paradoxes of time travel, which are well known to any fan of science fiction, demand a solution if we are to take seriously the possibility of the existence of CTCs. After all, the world can't admit of a physical situation in which the actions of a time traveler prevent the creation of his own time machine. The classical proposal for solutions to the time travel paradoxes simply states that such a situation could not obtain because it is inconsistent. That is to say, the classical solution is to impose a global property of self-consistency on the events in spacetime in order to rule out the possibility of paradoxical situations arising (see e.g.\ \cite{Lewis1976} and \cite{Novikov2002}).

David Deutsch argued in his 1991 paper ``Quantum Mechanics Near Closed Timelike Lines" \cite{Deutsch1991} that, under certain assumptions, quantum mechanics can solve the paradoxes associated with time travel to the past. What bothered Deutsch about the classical solutions to these paradoxes was the element of superdeterminism implicit in them. Certain initial states of systems are ruled out by these classical solutions, in order to preserve a global consistency. This is at odds with what Deutsch identifies as one of the fundamental principles of the philosophy of science: that global constraints should not overrule our ability to act locally in accord with the laws of physics. He calls this the \emph{autonomy principle}.  The classical consistency condition violates this principle by disallowing certain initial trajectories of systems traveling along CTCs.

Deutsch showed that taking quantum effects into account allowed for a solution to the paradoxes of time travel, without disallowing any initial states of the system. He showed that for any initial condition, there is a quantum fixed point solution representing a self-consistent physical state of the system. This is achieved by allowing for mixed quantum states to obtain on the CTC---a strategy to which solutions in the classical setting do not have access. These mixed states represent, for Deutsch, a collection of worlds, or ``multiverse".

The Deutsch closed timelike curve (D-CTC) model has been influential in the quantum foundations literature as a plausible candidate for how negative time delays would work in terms of information flow (see e.g.\ \cite{BrunEtAl2009}, and more recently \cite{RingbauerEtAl2014} and \cite{BubStairs2014}). The operational features of the model are taken on board, and are considered to be unproblematic additions to the machinery of the quantum information approach.

Presumably, the justification in doing this comes from the assumption that the multiverse on which Deutsch is relying in his description of the D-CTC model is the ``multiverse" of the Everett interpretation of quantum mechanics. If this were true, it could safely be ignored by those preferring an operationalist version of quantum theory, since the Everett interpretation is, after all, unmodified quantum mechanics.

However, Deutsch is relying on the existence of a more general notion of the multiverse, wherein the universes are not generated as the result of the Schr\"odinger evolution of the universal wavefunction, leading to the branching-off of macroscopic worlds, as in the standard Everett picture. Rather, the individual universes in this case exist timelessly and in parallel, many identical with one another for at least some period of time. These are not the many worlds of the Everett interpretation.

Part of the reason that this is confusing is that Deutsch refers to both of these objects by the term ``multiverse". For my purposes, I will refer to the collective many worlds of the standard Everett interpretation as the \emph{Many-Worlds Multiverse} (MWM). These are the result of branching of the universal wavefunction via decoherence. I'll refer to this other multiverse concept as the \emph{Mixed-State Multiverse} (MSM). The reason for this will become clear. These universes are a kind of parallelism of existent worlds. They are not generated by the evolution of the quantum state of the universe. They timelessly exist in parallel with one another.

I will argue that a close analysis of the details of Deutsch's model shows that it cannot be so easily separated from his deep metaphysical commitments to the real existence of parallel universes. These parallel worlds are importantly different from the many worlds of the standard Everett interpretation, and as such, Deutsch's key structure is not supported by quantum theory.

Finally, I'll address the following question: Is it still possible to adopt the purely operational features of his model? I'll argue that Deutsch uses the existence of MSM in the reasoning about the operation of the model. I'll show, by considering a simple example, that a purely operational acceptance of the D-CTC model would allow for predictions that Deutsch explicitly rules out. That is to say, Deutsch relies on features of the implicit underlying metaphysical picture when defining the effects of his model, and without this influence, different predictions are possible.

%%%%SECTION 2

\section{Deutsch's CTC Model}

In \cite{Deutsch1991}, Deutsch introduces a model for the analysis of the physical behavior of CTCs.  Prior to his work, the standard way of analyzing the physical effects of chronology-violating regions of spacetime was in terms of their underlying geometry. Deutsch considered this approach to be insufficient because it fails to take quantum mechanical effects into account.  He proposed an alternative approach which involves analyzing the behavior of CTCs in terms of their information processing capabilities, by redescribing the information flow of the physical situation in the form of a quantum computational circuit.  

His proposal is to introduce a simple standard form into which any spacetime-bounded network containing a CTC can be trivially transformed for the purpose of analysis.  The simple standard form involves translating all spacetime-bounded networks into circuits in which each particle traveling in the original network is replaced by sufficiently many carrier qubits to encode the information flow.  The regions in which the particles interact are localized (by denotationally trivial transformations) into gates, such that the states of the particles do not change while traveling between them.  And finally, all chronology-violating effects of the network are localized to sufficiently many carrier particles on closed loops, which only interact with chronology-respecting particles in gates.

Deutsch points out that chronology violation itself makes no difference to the behavior of a network unless there is a closed loop of information.  In the original network, this closed information path could potentially not be confined to the trajectory of any single particle (since the carriers can interact with each other), but for any such network, there is a denotationally trivial transformation which will localize the closed information path on sufficiently many carriers on closed paths.

In the classical case, networks containing chronology violations can lead to paradoxes that seem to put unnaturally strong constraints on possible initial conditions of physical systems (e.g.\ you are somehow prohibited from getting in the time machine that would take you back to kill your grandfather).  Deutsch uses his model to argue that, when quantum mechanics is taken into account, these unnatural constraints on initial states disappear.  Deutsch's fixed point theorem states that CTCs ``place no retrospective constraints on the state of a quantum system" \cite{Deutsch1991}.  That is to say, for any possible input state, there will be a paradox-free solution.

This is the result of a consistency condition implied by the quantum mechanical treatment of time-traveling carrier particles interacting with later versions of themselves.  If we let \(\left|\psi\right>\) be the initial state of the ``younger" version of the carrier particle, and let \(\hat{\rho}\) be the density operator of the ``older" version of the carrier particle, then the joint density operator of the two particles entering the region of interaction is \[\left|\psi\right>\left<\psi\right|\otimes\hat{\rho}\] and the density operator of the two carrier particles after the interaction is \[U(\left|\psi\right>\left<\psi\right|\otimes\hat{\rho})U^{\dagger}\] where \emph{U} is the interaction unitary.  The consistency condition requires that the density operator of the younger version of the carrier particle as it leaves the region of interaction is the same as that of the older version as it enters the region of interaction.  \[\hat{\rho}=\textrm{Tr}[U(\left|\psi\right>\left<\psi\right|\otimes\hat{\rho})U^{\dagger}]\]  This makes intuitive sense, as it seems natural to describe the evolution in the following ``pseudo-time" narrative: the interaction causes the earlier version of the carrier particle to become the later version.\footnote{Pseudo-time narratives like this can be useful for thinking about the effects of a CTC, but cannot be taken literally, as the state of the system \(\hat{\rho}\) on the CTC was the same prior to the interaction.}  When translated via a denotationally trivial transformation to a network in which the chronology-violating behavior is localized to a single particle on a CTC that interacts with a chronology-respecting (CR) carrier particle, the consistency condition for the CTC system is \[\rho_{\textrm{\scriptsize CTC}}=\textrm{Tr}_{\textrm{sys}}[U(\left|\psi\right>\left<\psi\right|\otimes\rho_{{\textrm{\scriptsize CTC}}})U^{\dagger}].\]  This requirement says that the density operator of the system on the CTC \emph{after} the interaction is the same as it was \emph{before} the interaction.  That is to say, after the interaction, the carrier particle on the CTC enters the ``future mouth" of the wormhole, and exits the ``past mouth" of the wormhole \emph{before} the interaction.  The state of the particle that comes out of the past mouth must be the same as the system that enters the future mouth.  Furthermore, \(\rho_{\textrm{\scriptsize CTC}}\) depends on \(\left|\psi\right>\), so the input state on the CR carrier particle has an effect on the state of the particle with which it will interact.

The puzzle arises, though, when we note that the CTC qubit must \emph{always} have been in this state.  There are no previous interactions with the CR qubit to force \(\hat{\rho}\) to evolve over time into the state that guarantees consistency.  Although Deutsch's model has avoided the superdeterminism of the classical solution to the time travel paradoxes, which constrained the initial states of the CR system, it seems to have introduced significant constraints in another place. Something like Lewis's classical consistency condition (see \cite{Lewis1976}) must still be at play. That is to say, there must be a deeper metaphysical justification (i.e.\ the impossibility of a self-contradictory history) which is behind Deutsch's quantum condition.

%%%%SECTION 3

\section{The Problem}

Deutsch's example of a Grandfather Paradox circuit clearly shows how this effect works. He models the information flow in the way shown in Figure 1.\begin{figure}[h!]
\centering
\caption{\footnotesize{The classical grandfather paradox circuit. From \cite{Deutsch1991}.}} 
\includegraphics[width=.45\textwidth]{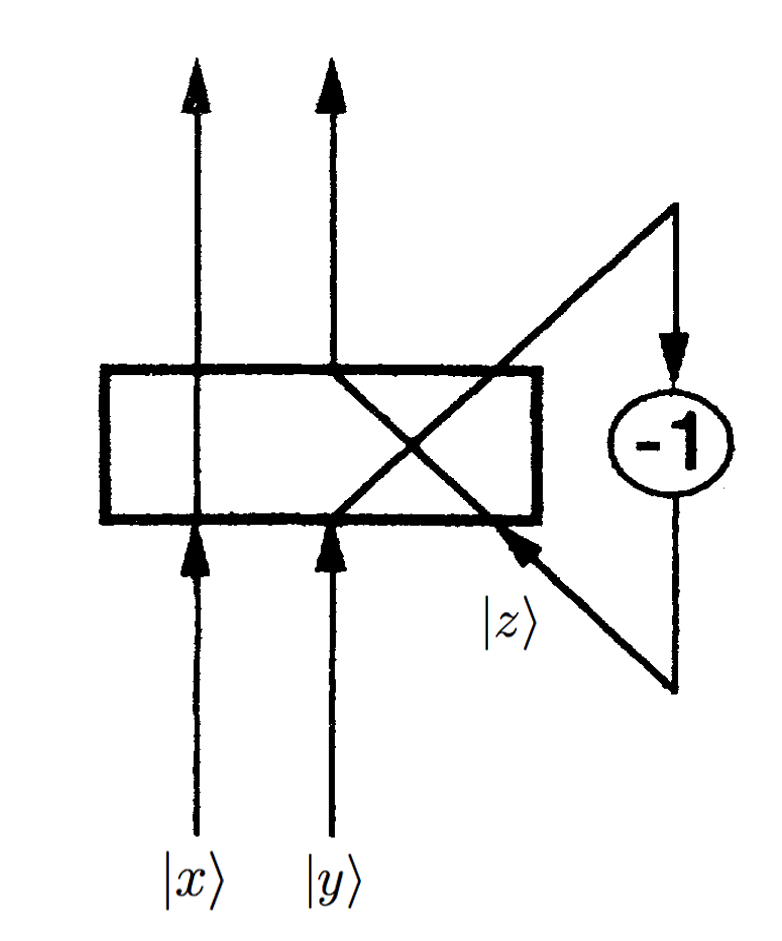}
\end{figure} The input states allowed in this setup are \(\left|1\right>_x\left|0\right>_y  \) and \(\left|0\right>_x\left|1\right>_y  \). The former represents an initial state with a time traveler present on a trajectory that \emph{does not} take her through the CTC, and the latter represents an initial state with a time traveler on the trajectory that \emph{does} take her through the CTC. That is, when \(x=1\) and \(y=0\), no time travel occurs. But when \(x=0\) and \(y=1\), a grandfather paradox becomes a possibility.

There are three inputs to the rectangular region of interaction, \(\left|x\right>\), \(\left|y\right>\), and the older version of the second qubit after it has traversed the CTC, which will be referred to as \(\left|z\right>\). Since \(\left|z\right>\) is simply an older version of \(\left|y\right>\) after they interact, and by stipulation there is no evolution of the state of the qubit on the CTC, the post-interaction state of \(\left|y\right>\) must equal the pre-interaction state of \(\left|z\right>\). This is a statement of Deutsch's consistency condition.

In the classical version of this circuit, the particles must be in pure states. The state of \(\left|z\right>  \) is equal to \(x \oplus 1\), because all that is being represented in this model is the presence of a bit on that channel. If \(x=1\), then no bit will go around the CTC. If \(x=0\), then there is a bit present on the CTC.

To enact the grandfather paradox, Deutsch sets the interaction in the region to be: \[ \left|x\right>_x\left|y\right>_y\left|x\oplus 1\right>_z \Rightarrow \left|x \oplus x \oplus 1\right>_x\left|y \oplus x \oplus 1\right>_y\left| x \oplus 1\right>_z. \] As noted above, the post-interaction state of \(\left|y\right>  \) must be equal to the pre-interaction state of \(\left|z\right>  \). That is, the following equivalence must hold: \[y \oplus x \oplus 1 = x \oplus 1.\] In the classical case, this can only be true if \(x=1\) and \(y=0\), meaning that no time traveler ever went back to prevent herself from traveling. The input state \(x=0\) and \(y=1\) would not yield a consistent solution, and is therefore ruled out. That would represent a situation where the time traveler succeeded in going back in time and prevented herself from entering the future mouth of the CTC. This paradoxical situation can not obtain, so the classical consistency condition rules out this input state.

The quantum solution must be formulated in terms of the fully transformed circuit---something Deutsch refrains from doing explicitly in \cite{Deutsch1991}. However, when the quantum solution is clearly diagrammed, the seeds of the problem for Detusch's account becomes apparent.\begin{figure}[h!]
\centering
\caption{\footnotesize{The fully-transformed grandfather paradox circuit. Two CNOT interactions, followed by a SWAP.}} 
\includegraphics[width=.75\textwidth]{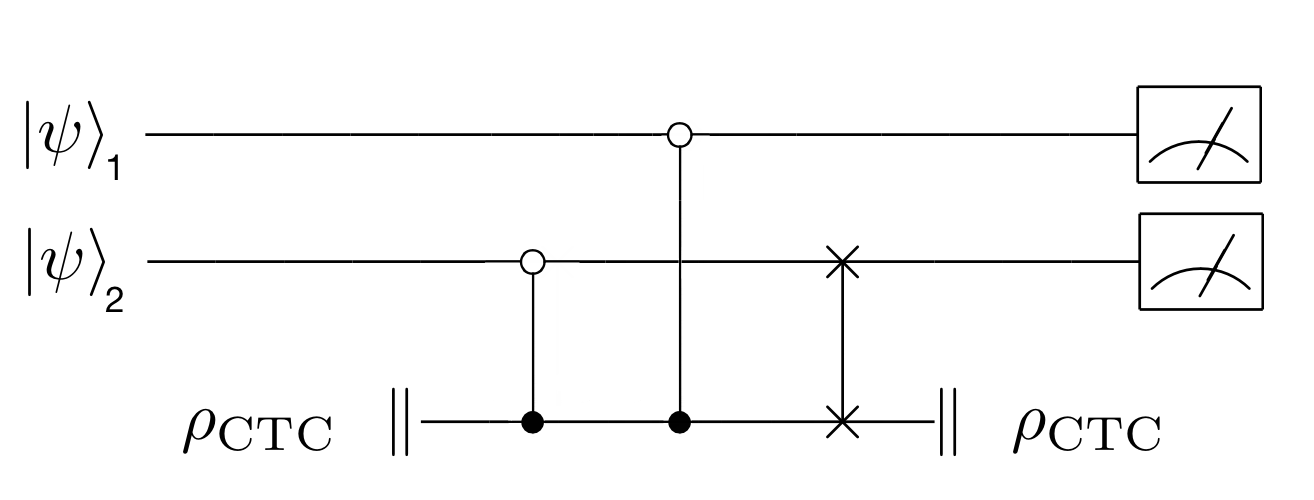}
\end{figure} The first two interactions from the CR perspective are ``controlled--NOT" gates. That means that if the controlling qubit (in this case, \(\rho_{{\textrm{\scriptsize CTC}}}\)) is in the state \(\left|1\right>\), then it will effect a NOT transformation on the target system.\footnote{NOT is the ``bitflip" operation, which takes \(\left|0\right>\rightarrow\left|1\right>\) and \(\left|1\right>\rightarrow\left|0\right>\). This is equivalent to the interaction from Deutsch's example, where the value of bit \(z\) is added to \(x\) and to \(y\).} The third interaction is SWAP, where the state of the system \(\left|\psi\right>_2\) is exchanged for the state of \(\rho_{{\textrm{\scriptsize CTC}}}\). The information flow here is identical to the partially-transformed circuit of Deutsch's example. However, whereas the state of the bit \(\left|z\right>  \) from the first circuit is clearly dependent on the state of \(\left|x\right>  \) (e.g.\ if \(x=1\), then \(y=0\), and there would be no bit to travel around the CTC), in the fully-transformed version the state of \(\rho_{{\textrm{\scriptsize CTC}}}\) is clearly \emph{uncaused}. The independence of the state of the system on confined to the CTC is more obvious in this version of the circuit diagram.

It is this feature of the fully-transformed circuit that Deutsch relies on for the quantum solution to the paradox. From the CR perspective, \(\rho_{{\textrm{\scriptsize CTC}}}\) has always been in the particular mixed state that allows for the success of the circut. When \(\rho_{{\textrm{\scriptsize CTC}}}=\frac{1}{2}\hat{I}\), then there will be a self-consistent solution for the classically forbidden input state \(\left|0\right>_1\left|1\right>_2\). This will yield the following output: \[\hat{\rho}= \frac{1}{2}\left( \left|00\right> \left<00\right| + \left|11\right> \left<11\right|   \right).  \] This mixed output state is interpreted by Deutsch as indicating that there are many separate worlds in which the two possible outcomes happen. The solution relies on the existence of parallel worlds into which a time traveling system goes when passing though a CTC. \begin{quote} The key thing to bear in mind when trying to visualize it is that in half of the universes (let us call them the ``\(A\) universes") the encounter happens and in the other half (the ``\(B\) universes") it does not happen. [...] In the \(A\) universes an observer appears ``from nowhere" (no one having embarked on a chronology-violating trajectory in that universe) and in the \(B\) universes an observer enters and disappears ``into nowhere" (since no one has emerged on the chronology-violating trajectory in that universe). But of course it is not really ``from nowhere" and ``into nowhere," but from and into other universes. \cite{Deutsch1991}  \end{quote}

For Deutsch, the mixed state \(\frac{1}{2}\hat{I}\) represents the connection between parallel worlds. In half of them the state of the system on the CTC is \(\left|0\right>\), and in the other half the state is \(\left|1\right>\). They travel across these bridges, into parallel worlds and interact with the CR systems there. The overall ensemble of states in these separate universes is the object to which the consistency condition applies. \begin{quote} In the Everett interpretation it is only the state, which describes, roughly speaking, a collection of values taken as a whole, which must be unchanged after passage round a closed timelike line.  \cite{Deutsch1991} \end{quote} The mixed state representing the collection of worlds must be self-consistent, but as we saw from the above example, the actual outputs of the circuit will be different in different universes, meaning in some the state that the CR qubits interact with is \(\left|0\right>\), and in the others it is \(\left|1\right>\).

%%%%%SECTION 4

\section{MWM vs.\ MSM}

Deutsch's model is often adopted wholesale by theorists working in the foundations of quantum theory. Deutsch's talk of travel between parallel worlds is thought to be unproblematic because it is assumed that he is simply adopting the language of the Everett interpretation of quantum theory. 

But there is an immediate problem with this interpretation of Deutsch's model. Imagine a time traveler traveling from \(t=2\) back in time to \(t=1\). She can't be traveling into her own past, because her presence there would  change it, leading to a different future evolution of the wavefunction, undermining the existence of the branch from which she came. She needs to travel to an already existent branch with an identical copy of herself at \(t=1\). The problem is, according to the Everett interpretation, there would be no such branch. Since the state of the world at \(t=1\) in the time traveler's actual past is, by stipulation, identical to the state of the world at \(t=1\) in the universe into which she is traveling, there would never have been a branching event that would have created multiple copies of the world.\footnote{The would, of course, be a multiplicity of branches on which the state of the world is slightly different from the state in the time traveler's past, but Deutsch requires that time travel be between worlds with \emph{identical} histories.} The existence of this destination world is not consistent with the branching structure of the standard Everett interpretation. Deutsch cannot be relying on the structure of MWM for his solution to the paradoxes of time travel.

However, this logical problem is obscured by the fact that Deutsch often talks about the Everett interpretation in terms of parallel existent worlds (e.g.\ see  \cite{Deutsch1997}). Deutsch admits that this talk of parallelism represents nothing more than a convenient approximation to the Everett interpretation, to be used with care. He states this explicitly and often: \begin{quote} The idea that quantum theory is a true description of physical reality led Everett and many subsequent investigators to explain quantum-mechanical phenomena in terms of the simultaneous existence of parallel universes or histories. [...] However, if reality---which in this context is called the \emph{multiverse}---is indeed literally quantum-mechanical, then it must have a great deal more structure than merely a collection of entities each resembling the universe of classical physics. For one thing, elements of such a collection would indeed be `parallel': they would have no effect on each other, and would therefore not exhibit quantum interference. For another, a `universe' is a global construct---say, the whole of space and its contents at a given time---\emph{but since quantum interactions are local, it must in the first instance be local physical systems, such as qubits, measuring instruments and observers, that are split into multiple copies, and this multiplicity must propagate across the multiverse at subluminal speeds.} \cite[emphasis added]{Deutsch2002} \end{quote}

If the D-CTC model cannot be embedded in the standard branching MWM, how then is it supposed to work? A close reading of Deutsch reveals that he has a very particular metaphysical view: He is a realist about the existence of many (possibly infinitely many) parallel worlds, which differ from MWM. That is to say, in addition to the many worlds that exist in the Everett interpretation as the result of the evolution of the wavefunction, he also is committed to the existence of a multiverse whose worlds exist eternally, and any two of which may be identical to one another for some or all of their histories.

The D-CTC model is crucially grounded in his commitment to this metaphysical view. The density matrices that represent the mixtures on the CTC are, for Deutsch, a collection of parallel worlds. But these worlds could not exist as the result of branching of the wavefunction. Part of the difficulty in determining his commitments comes from the fact that Deutsch has presented his solutions to the paradoxes of time travel in popular works much more frequently than in scholarly works. But I believe a clear picture of what he has in mind can still be developed.

It's clear from his popular discussions of his D-CTC model that he considers a kind of parallelism to be conceptually foundational to his solutions to the paradoxes of time travel (see \cite{Deutsch1994} and \cite[Ch.\ 12]{Deutsch1997}). The grandfather paradox is solved (while preserving the autonomy principle), because time travel takes us into a another universe. A time traveler is free to kill the person she meets there (a counterpart of her own grandfather), because she will not actually be altering her own past. Rather, she is participating in the past of another universe.

In these writings, he claims to ground the existence of these parallel worlds in the Everett interpretation. \begin{quote} According to Everett, if something physically can happen, it does---in some universe. Physical reality consists of a collection of universes, sometimes called a multiverse. [...] What then, does quantum mechanics, by Everett's interpretation, say about time travel paradoxes? Well, the grandfather paradox, for one, does not arise. [...] If the classical space-time contains CTCs, then, according to quantum mechanics, the universes in the multiverse must be linked up in an unusual way. Instead of having many disjoint, parallel universes, each containing CTCs, we have in effect a single, convoluted space-time consisting of many connected universes. \cite{Deutsch1994} \end{quote} However, for the reasons discussed above, the universes he describes cannot be the result of branching. Universes only branch off from one another when there are significant enough differences between them such that they would lead to macroscopically distinguishable worlds. As he says in \cite{Deutsch2010}, Everett worlds are emergent, and result from the process of decoherence. \begin{quote} ...Only [the Everett interpretation] can accommodate the fact that universes turn out to be approximate, emergent structures in the multiverse.\footnote{When discussing the Everett interpretation, Deutsch's terminology varies from the standard. As he says in \cite{Trans} ``I'm quite happy to call the multiverse the universe and the universe a branch." This sentence should be understood to mean that branches emerge via decoherence, and are elements of the universal wavefunction. This is consistent with Wallace's presentation in \cite{Wallace2012}. } Decoherence theory opened up the study of the structure of the multiverse: not just how the quasiclassical universes emerge, but also how what is happening exactly when the universes are present emergently.  \cite{Deutsch2010} \end{quote} But compare that to the statement of the existence of the parallel worlds he makes in \cite{Deutsch1994}: \begin{quote} If the classical space-time contains CTCs, then, according to quantum mechanics, the universes in the multiverse must be linked up in an unusual way. Instead of having many disjoint, parallel universes, each containing CTCs, we have in effect a single, convoluted space-time consisting of many connected universes. The links force [the time traveler] to travel to a universe that is identical, up to the instant of her arrival, with the one she left, but that is thereafter different because of her presence. \cite{Deutsch1994} \end{quote} And another statement from the conclusion of the original article: \begin{quote} I have shown that the traditional ``paradoxes'' of chronology violation, whatever position one takes on their seriousness, do not occur at all under quantum mechanics. \cite{Deutsch1991}  \end{quote}

Deutsch considers it a lesson of this analysis that there are multiple parallel worlds connected up in the right kind of way. \begin{quote} So, for time travel to be physically possible it is necessary for there to be a multiverse. And it is necessary that the physical laws governing the multiverse be such that, in the presence of a time machine and potential time travellers, the universes become interconnected in the way I have described, and not in any other way. \cite{Deutsch1997} \end{quote} But these are not the many worlds of the branching Everett model. They are from some larger multiverse. \begin{quote} What is needed is to express such arguments in the framework of a theory of a multiverse---sometimes it has to be a bigger multiverse than Everett's.  \cite{Deutsch2010} \end{quote} What grounds this talk of parallelism? The root of the idea has to do with the fact that the state \(\rho_{{\textrm{\scriptsize CTC}}}\) is mixed in the solution to the paradox. This mixed state does not arise as the result of the evolution of the system from its initial conditions. Rather, it is induced by the presence of the CTC. This feature of the view will be discussed below in Section 6. First, though, I will address an argument from David Wallace that the MSM is the necessary consequence of any quantum model of chronology violation which allows for entangled particles to enter a CTC.

%%%%SECTION 5

\section{Wallace's Analysis}

David Wallace, in \emph{The Emergent Multiverse}, addresses some of these same issues with the D-CTC model. He says \begin{quote} ...[T]he multiplicity present in a mixed state is, as we have seen previously, of a rather different kind from conventional Everettian multiplicity. The parallel universes of the Everett interpretation are present even in pure states, are emergent phenomena whose concrete form is given by decoherence theory, and affect one another dynamically via interference; the multiple `Universes' (actually, multiple entities which are themselves multiplicities of quasi-classical branches) in mixed states are present by interpretive fiat. \cite{Wallace2012}\end{quote}  He agrees that what Deutsch manages to establish is that introducing parallel worlds can solve the paradoxes of time travel, and not that these worlds follow from the Everett interpretation itself. \begin{quote} So far, then, all I think we've established (and all Deutsch establishes) is that accepting parallel universes can solve time travel's contradiction paradoxes. There is nothing quantum-mechanical about this observation, and it would be reasonable to regard the quantum and time-travel parallel universes as being of an entirely different kind. \cite{Wallace2012} \end{quote}

However, Wallace argues that this conclusion is too quick, and provides a simple argument meant to show that the Deutsch MSM is the inevitable result of any model that would allow for time travel of systems that are entangled with systems in the CR region to any degree.

Wallace wants to show that any model for time travel of a quantum system, as long as entangled systems can travel in time, would necessitate that a mixed state be present on the CTC. He asks us to consider a single particle traveling through a CTC into the past, not interacting with itself in any way, and traveling on into the unambiguous future unchanged. Clearly there would be no contradiction entailed by this physical situation, so there is no need to worry about what the state bound to the CTC would be in the transformed version of the circuit. If we consider a denotationally trivial transformation in which there is a qubit in the state \(\rho_{\textrm{\scriptsize CTC}}\) bound to the CTC , on which a SWAP is performed with the input system, the consistency condition would entail that the state on the CTC was equal to the state of the system entering the region of interaction from the unambiguous past. If the system entering the region of interaction was in the state \(\left|1\right>\), then \(\rho_{\textrm{\scriptsize CTC}}=\left|1\right>\left<1\right| \).

Now Wallace argues that if the state of the input system is mixed because it is entangled with another particle, then the same reasoning leads us to conclude that the state of the system bound to the CTC must also be mixed. Consider two external systems---the input system and an ancilla---in the nonseparable state \[ \frac{1}{\sqrt{2}}\left(\left |0\right> \otimes \left|0 \right> + \left|1 \right> \otimes \left|1 \right>    \right).  \] The reduced density matrix of each individual particle is \[\frac{1}{2}\left(\left |0\right>  \left<0 \right| + \left|1 \right>\left<1 \right|    \right). \] Therefore, Wallace argues, when one of these particles enters a chronology violating region, the state bound to the CTC must also be mixed in order to conform with the consistency condition.

\begin{figure}[h!]
\centering
\caption{\footnotesize{The Wallace example. In the case where \(\left|\psi\right>_1\) and \(\left|\psi\right>_2\) are entangled, he claims there must be a mixed state on the CTC.}} 
\includegraphics[width=.75\textwidth]{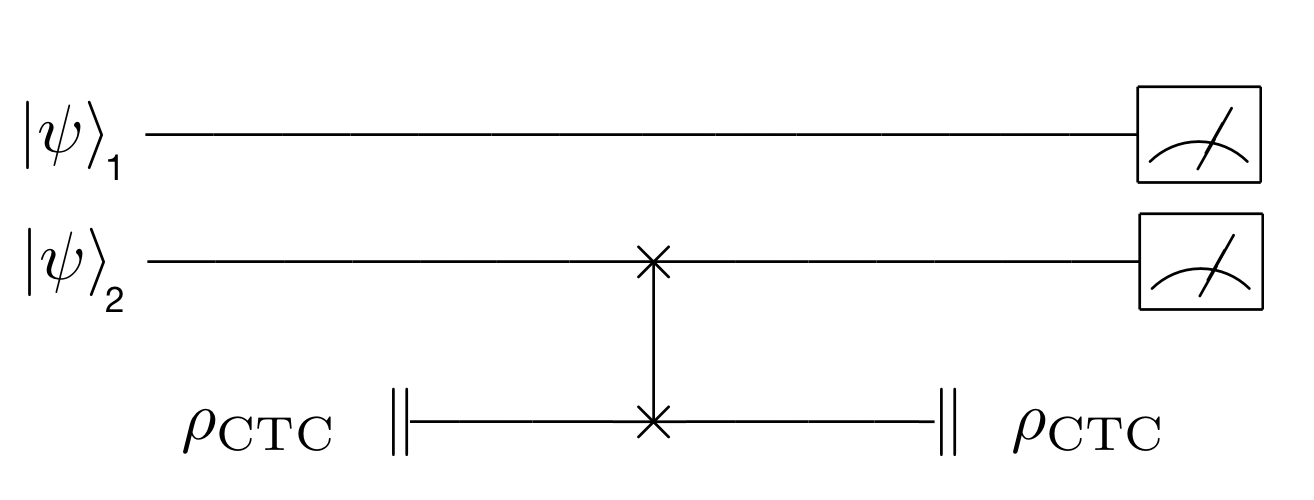}
\end{figure}

He goes on to conclude \begin{quote}  There is nothing particularly special about my example: essentially any situation in which part but not all of an entangled system undergoes time travel will require mixed states to satisfy the consistency condition. So either (i) in the presence of closed timelike curves the Universe can be in a genuinely mixed state, or (ii) entangled states cannot travel on closed timelike curves. \cite{Wallace2012}  \end{quote}

However, Wallace's example, and hence the argument as a whole, is based on a misreading of the D-CTC model. A D-CTC circuit is not a faithful replication of the topology of the physical situation being modeled. Rather, each D-CTC circuit is a representation of the information flow of the physical situation. In this section I will argue that, in Wallace's example, the negative time delay is irrelevant to the information flow, and therefore would not be represented in the D-CTC circuit designed to model the physical situation he uses as his example. Deutsch's consistency condition only applies to the state of the carrier qubit on the fully-transformed simple standard-form circuit. Since, in the model of the physical situation Wallace describes, there is no state bound to the CTC (as there is no closed path of information to represent), the consistency condition does not apply.

To begin, note that Deutsch characterizes his CTC model as ``representative": \begin{quote} The basic method of this paper is to regard computations as representative physical processes---representing the behavior of general physical systems under the unfamiliar circumstances of chronology violation. \cite{Deutsch1991}  \end{quote} His standard form for the D-CTC circuits is based on the idea of the denotationally trivial transformation, where denotational equivalence is understood in terms of mapping of inputs to outputs. \begin{quote} Insofar as we wish to study chronology--violating regions solely in terms of their external responses to external stimuli [...] we are entirely free to use denotationally trivial transformations to simplify networks that model the effects of the region. \cite{Deutsch1991}\end{quote} The physical effects of the spacetime that contains a chronology--violating region are encoded in the information--theoretic features of a representative circuit. That circuit is transformed into a simple form via denotationally trivial transformations, such that the chronology--violating information flow is localized in the unitary interactions between qubits bound to the CTC and qubits in the CR region. He goes on to say \begin{quote}  But it must be borne in mind that as regards what happens \emph{inside} a chronology-violating region, a computational network that faithfully models the physics of the region may cease to do so under certain denotationally trivial transformations. Therefore when transforming networks one should where appropriate take note of what the transformations correspond to in the spacetime that is being modeled.   \cite{Deutsch1991}\end{quote} The important point here is that the physical situation of chronology violation is being represented by a circuit in terms of the information flow of representative carrier particles. The arrangement of the components of the circuit may look very different from the physical situation being modeled.

Most importantly for this discussion of Wallace's example is a point Deutsch makes about when the D-CTC model (and therefore his consistency condition) applies to a physical situation. \begin{quote}Chronology violation in itself (i.e., the existence of negative delays) makes no fundamental difference to the behaviour of a network unless there is a \emph{closed path for information}. [...] If a network has no closed path for information there is always a denotationally trivial transformation that removes all negative delays by introducing positive delays elsewhere. If the network does have such a path then there is always a denotationally trivial transformation that makes all negative delays occur on \emph{closed} trajectories of carriers. \cite{Deutsch1991}\end{quote} And later: \begin{quote}  All the special properties of chronology-violating networks are consequences of consistency conditions around those closed trajectories. \cite{Deutsch1991} \end{quote} A physical situation involving chronology violation is modeled by first assessing its information flow. Encode the relevant physical features in the states of representative qubits. Then identify the closed path of information. Perform whatever denotationally trivial transformation is necessary to localize the closed path of information on a closed trajectory of as many qubits as necessary. This is \(\rho_{\textrm{\scriptsize CTC}}\). It is to this object that the consistency condition applies.

The problem with Wallace's example is that he conflates the representation (the standard form D-CTC circuit) with the physical situation being described. He asks us to imagine a standard--form D-CTC circuit with a unitary interaction (a single SWAP) which entails that there is no closed path for information. Since there is no interaction between the input qubit and \(\rho_{\textrm{\scriptsize CTC}}\) that has any effect on the output, there is a denotationally trivial transformation that would redescribe the physical situation without the inclusion of any negative time delays. 

In other words, Wallace is asking us to imagine a physical situation where a single particle enters the future mouth of a wormhole, is sent back in time, exits the past mouth, and does not interact with the younger version of itself. Since there is no self-interaction, the state of the system undergoing this evolution is irrelevant to the description. In a situation like this, it is not necessary for other-worldly counterparts to prevent a paradox, so the evolution of the system through the chronology-violating region can take place within a single universe.

Interestingly, Wallace takes his example to show that D-CTCs necessarily break entanglement. However, according to what has just been argued, it is possible for a system to travel through a region of chronology violation and preserve its entanglement with systems in the CR region. The D-CTC model doesn't entail that entanglement is always broken in cases of chronology violation. But in many cases the physical situation will be more complex than the example currently under consideration. According to Deutsch, the connection between parallel MSM worlds only occurs when there is a closed loop of information. If a physical system travels back in time on such a trajectory that it will interfere with itself, the D-CTC model says that it will actually pass into another universe. A complex macroscopic physical system will almost certainly have degrees of freedom that are entangled with systems that remain in the CR region. That entanglement will indeed be broken when the system of interest passes into another world via the CTC (the system exiting the past mouth of the CTC will come \emph{from} another universe, and therefore cannot be entangled with the CR systems in the original universe). But in cases like the one Wallace describes, there is no closed path of information, and therefore no negative time delay in the transformed version of the D-CTC representation of the situation. The consistency condition does not apply, and therefore the particle that travels around the CTC remains in its own universe, and its entanglement with CR systems remains.

Wallace's discussion of the D-CTC model is mostly correct, but the argument in favor of the idea that MSM parallelism is an inevitable consequence of any quantum theory of chronology violation that allows for systems in entangled states to act as inputs to CTC is not convincing. In addition, there are alternative models for quantum CTCs, most notably the P-CTC model (see, e.g., \cite{LloydEtAl2010} and \cite{LloydEtAl2010b}) that explicitly allow for entangled systems to travel back in time while preserving their entanglement. These models can be embedded in a single universe, and the explanations they give do not need to be supplemented by the MSM.

%%%%SECTION 6

\section{Mixed States Bound to the CTC}

The power of the D-CTC model is that it allows for the information flow bound to the CTC to be in a mixed state, which can solve the classically paradoxical situations. For example, in the Grandfather Paradox circuit, it is because the closed loop of information bound to the CTC is free to be in the state \(\frac{1}{2}\hat{I}\) when the input state is \( \left|0 \right>_1 \left|1 \right>_2\) that there is a self-consistent solution.

But where do these mixed states come from? It might be argued that the reason the state exiting the past mouth of the CTC in Figure 2 is \(\frac{1}{2}\hat{I}\) is because the interaction that \(\left|\psi\right>_2\) underwent \emph{caused} it to be so. But this reasoning implicitly relies on a pseudotime narrative of the interaction. In addition, Deutsch explicitly denies that the mixed state should be interpreted as a statistical ensemble. And in reality, there is no beginning to the existence of the state bound to the CTC. It is a closed loop of information, and as such can have no external cause. It must therefore be a feature of the existence of the CTCs that allow those systems to evolve from pure states into mixed states.  

These states \emph{simply exist}, present on the CTC, not caused by any interaction, and therefore not explainable. This is another way of stating the additional element of superdeterminism present in his model---it is this structure that he takes to ground his claim that the universes simply ``connect up" in the right way.

Since the state on the CTC isn't caused by the interaction with the initial state, but merely exists because it allows for a consistent solution, this gives rise to the possibility of an alternative prediction in the case of the classically allowed input state. Deutsch gives only an extra-theoretic answer to why the following solution is ruled out. It is not inconsistent with the consistency condition.

When describing the classical version of the Grandfather Paradox circuit (Figure 1), Deutsch defines the state of the bit \( \left|z\right>\) as \(\left| x \oplus 1   \right>  \). That is to say, if we start with the time traveler present on the trajectory that avoids the CTC (represented by \(x=1\), \(y=0\)), then there won't be anything to travel around it, and if the time traveler starts on the trajectory that will take her through the CTC (represented by \(x=0\), \(y=1\)), then the CTC will be occupied. So the state on the CTC should be the opposite of \(\left| x   \right>  \).

But in this fully transformed version of the circuit (Figure 2), the independence of the state of the system on the CTC (representing the closed loop of information) is more obvious. The only condition is that the state of \(\rho_{\textrm{\scriptsize CTC}}\) when it enters the future mouth of the CTC is the same as the state of \(\rho_{\textrm{\scriptsize CTC}}\) as it exits the past mouth of the CTC.

Since there is no way to account for why \(\rho_{\textrm{\scriptsize CTC}}\) takes on the particular value it does when exiting the past mouth of the CTC in the cases where it solves the paradox---after all, it is uncaused---then there is likewise no way to account for why \(\rho_{\textrm{\scriptsize CTC}}\) has the state it does when exiting the past mouth of the CTC when its state is irrelevant to a paradox. The only condition in the former case is that its state is consistent with \( \rho_{ \textrm{\scriptsize CTC}}\) entering the future mouth. It seems that this should be the same in the latter case.

If that's true, then there is another consistent solution to the case where the initial state of the system is \( \left|1\right>_x \left|0\right>_y\), namely the one that takes the final state to \( \left|0\right>_x \left|1\right>_y\). If the system  \(\rho_{\textrm{\scriptsize CTC}}\) exits the past mouth of the CTC in the state \( \left|1\right>_z\), then the effect of the two CNOT gates will evolve the system in the following way \( \left|1\right>_x \left|0\right>_y  \Rightarrow \left|0\right>_x \left|1\right>_y\), and then the SWAP gate will have no effect.

In the notation of the example from \cite{Deutsch1991}, we are denying that the state of  \( \left|z\right>\) should equal \(\left| x \oplus 1   \right>  \). Since  \( \left|z\right>\) can take on any state, there are two solutions to the input state  \( \left|1\right>_x \left|0\right>_y\). Recall, the interaction was defined as:  \[ \left|x\right>_x\left|y\right>_y\left|x\oplus 1\right>_z \Rightarrow \left|x \oplus x \oplus 1\right>_x\left|y \oplus x \oplus 1\right>_y\left| x \oplus 1\right>_z \] which, with the modification under consideration becomes:
 \[ \left|x\right>_x\left|y\right>_y\left|z\right>_z \Rightarrow \left|x \oplus z \right>_x\left|y \oplus z \right>_y\left| z \right>_z \] The consistency condition requires that the pre-interaction state of \( \left|z\right>\) equals the post-interaction state of \( \left|y\right>\).
 
 The standard solution for the input \( \left|1\right>_x \left|0\right>_y\), where \(z=0\) still obviously holds. But there is also a consistent solution for when \(z=1\): \[ \left|1\right>_x\left|0\right>_y\left|1\right>_z \Rightarrow \left|1 \oplus 1 \right>_x\left|0 \oplus 1 \right>_y\left| 1 \right>_z = \left|0\right>_x\left|1\right>_y\left|1\right>_z  \]
 
Deutsch would argue that this is an unphysical solution. Since the CTC is not in use, in all the identical parallel universes (to which a CTC could potentially connect us), the CTC will also be unused. \begin{quote} For example, if I am not going to use a time machine come what may, then no time-travelling versions of me must appear in my snapshot\footnote{He is using the term ``snapshot" as an analogue for ``timeslice" in this more general multiverse framework where a unique time ordering of a foliation may not be possible.}; that is, no universes in which versions of me do use a time machine can become connected to my universe. If I am definitely going to use the time machine, then my universe must become connected to another universe in which I also definitely use it. And if I am going to try to enact a `paradox' then, as we have seen, my universe must become connected with another one in which a copy of me has the same intention as I do, but by carrying out that intention ends up behaving differently from me. \cite{Deutsch1997} \end{quote} Therefore, the qubit that exists the CTC couldn't have come from anywhere. Deutsch says that the way the universes connect up depends on the intentions of the time traveler. \begin{quote}A real time machine, of course, would not face these problems. It would simply provide pathways along which I and my counterparts, who already existed, could meet, and it would constrain neither our behaviour nor our interactions when we did meet. The ways in which the pathways interconnect---that is, which snapshots the time machine would lead to---would be affected by my physical state, including my state of mind. That is no different from the usual situation, in which my physical state, as reflected in my propensity to behave in various ways, affects what happens. \cite{Deutsch1997} \end{quote} But this explanation requires the acceptance of the existence of the parallel universes of MSM. To attempt to adopt Deutsch's model without also bringing this sizable metaphysical commitment on board would leave you no way to rule out our alternative prediction.

%%%%SECTION 7

\section{Comments}

To accept the D-CTC model, it seems necessary to accept that the state of the universe is a density matrix with each elements corresponding to a physically real parallel world, which can, in the presence of a CTC, get connected up with our world, so that a collection of these separate universes all contribute elements to the ensemble of states that is present on the CTC, ensuring that interactions with the CR system will leave the ensemble in same state in which it began.

Deutsch includes a discussion of this possibility in a rather oblique way in \cite{Deutsch1991}. He argues that, since the existence of a CTC allows for a pure state to evolve into a mixed state, it may be worth reconsidering the question of whether it is possible that the whole universe should be described by a density matrix \(\hat{\rho}(t)\): \begin{quote} \(\hat{\rho}(t)\) therefore describes a collection of ``universes" (each one itself consisting of multiple universes under the Everett interpretation), one for each nonzero \(p_i\). Each evolves precisely as if the others were absent and it had a pure state \(\left|\psi_i (t)\right> \). This is quite unlike the Everett multivaluedness caused by the linear superposition of components of a state vector, which is detectable through interference phenomena. Thus the cosmology described by \(\hat{\rho}(t)\) contains a multiplicity of mutually disconnected and un-observable entities and is vulnerable to the ``Occam's razor" argument that is sometimes erroneously leveled against the Everett interpretation.

%%THESE TWO PARAGRAPHS STAY TOGETHER

But in the presence of closed timelike lines the evolution with respect to an external time coordinate is not longer necessarily unitary as in (38). [...] Therefore in principle it might be possible to detect experimentally the difference between distinct density operators with identical eigenstates, so the ``Occam's razor" argument no longer necessarily holds. \cite{Deutsch1991}  \end{quote} Of course, the existence of the mixed state on the CTC, which is required by Deutsch's consistency condition, is the \emph{reason} a pure state will evolve into a mixed state. And the interpretation of those mixed states presupposes that the universe is itself in a mixed state. Deutsch is using the claim that D-CTCs are supported by the Everett interpretation to bolster their credibility, and using their predictions to argue for the existence of MSM. But D-CTCs are only supported by Everett if you already accept that MSM is consistent with the Everett interpretation, and that the Everett worlds don't need to be the result of the Schr\"odinger evolution of the wavefunction and decoherence.

The majority of people writing on D-CTCs are interested in questions about quantum information and quantum computation, and generally think of the debates over the interpretation of quantum theory to be superfluous. To what degree could the operational feature of the D-CTC model be accepted without any commitment to the underlying metaphysics? 

The example of the alternative solution to the Grandfather Paradox circuit from the last section undermines the idea that D-CTCs can be operationally adopted without consequence. Deutsch would rule out the prediction, but his justification for this relies on his interpretation of the model, and not simply on the consistency condition.  He would argue that the solution is impossible because the universes aren't connected up in the right way, because the state on CTC is sensitive to the intentions of nearby experimentalists. This kind of explanation would not be formulable without reference to the existence of the parallel worlds. And without recourse to this explanation, it seems as though an operationalists who wish to adopt the D-CTC model will need to accept that makes  predictions different from Deutsch's in simple cases.

This analysis is by no means meant to undermine Deutsch's model---if you accept his assumptions, there is no problem. However, it is meant to challenge the claim that there is a quantum mechanical solution to the Grandfather Paradox. It is often claimed that the resources of the Everett interpretation (the existence of the Everett worlds) allow for a solution to the paradox. This is not so. Deutsch's solutions to the paradoxes of time travel are ingenious and fascinating, but they proceed from a substantive metaphysical commitment that takes them beyond the domain of pure quantum theory.

\bibliographystyle{abbrv}

\end{document}